# A LOW OVERHEAD MINIMUM PROCESS GLOBAL SNAPSHOT COLLECTION ALGORITHM FOR MOBILE DISTRIBUTED SYSTEMS

Surender Kumar [1],R.K. Chauhan[2] and Parveen Kumar[3]

[1]Deptt. of I.T, Haryana College of Tech. & Mgmt. Kaithal-136027(HR), INDIA
skjangra@hctmkaithal-edu.org
[2]Deptt. of Computer Sc & Application, Kurukshetra University, Kurukshetra(HR),India
rkcdcsa@gmail.com
[3]Department of Computer Sc. & Engineering, MIET, Meerut(U.P), India
Pk223475@gmail.com

### ABSTRACT

*Coordinated checkpointing is an effective fault tolerant technique in distributed system as it avoids the domino effect and require minimum storage requirement. Most of the earlier coordinated checkpoint algorithms block their computation during checkpointing and forces minimum-process or non-blocking but forces all nodes to takes checkpoint even though many of them may not be necessary or non-blocking minimum-process but takes useless checkpoints or reduced useless checkpoint but has higher synchronization message overhead or has high checkpoint request propagation time. Hence in mobile distributed systems there is a great need of minimizing the number of communication message and checkpointing overhead as it raise new issues such as mobility, low bandwidth of wireless channels, frequently disconnections, limited battery power and lack of reliable stable storage on mobile nodes. In this paper, we propose a minimum-process coordinated checkpointing algorithm for mobile distributed system where no useless checkpoints are taken, no blocking of processes takes place and enforces a minimum-number of processes to take checkpoints. Our algorithm imposes low memory and computation overheads on MH's and low communication overheads on wireless channels. It avoids awakening of an MH if it is not required to take its checkpoint and has reduced latency time as each process involved in a global checkpoint can forward its own decision directly to the checkpoint initiator.*

## 1. INTRODUCTION

Checkpointing is a well-established technique to deal with process failures and increase the system reliability and fault-tolerance in distributed systems [23]. In this approach, the state of each process in the system is periodically saved on stable storage, which is called a checkpoint of a process. To recover from a failure, the system restarts its execution from a previous error-free, consistent global state [3]. In a distributed system, since the processes in the system do not share memory, a global state of the system is defined as a set of local states, one from each process. The state of channels corresponding to a global state is the set of messages sent but not yet received. A global state is said to be "consistent" if it contains no orphan message; i.e., a message whose receive event is recorded, but its send event is lost [3]. A mobile system is a distributed system where some of processes are running on mobile hosts (MHs) [5].The term "mobile" means able to move while retaining its network connection. A host that can move while retaining its network connection is an MH. An MH communicates with other nodes of system via special nodes called mobile support station (MSS)[23].

In this paper, our main aim is to minimize the checkpointing overhead and reduces checkpointing latency. Our algorithms has not any useless checkpoints, forces only minimum of processes, reduced checkpoint latency (time between a process initiates a checkpoint request and global checkpointing process completes) and does not suspend their computation during checkpointing process. During normal message transmission, processes append the information regarding the set of processes which are directly or indirectly checkpoint dependent with the application message. By this way destination process updates its own dependency set and compute minimum set (set of processes which are directly or indirectly dependent). When a process initiates the checkpointing algorithm, it sends checkpoint request to all processes which belongs to minimum set simultaneously. After receiving the checkpointing request every involved node directly responds to the checkpoint initiator negatively or positively.

The rest of the paper is organized as follows. We formulate related work in 2, proposed checkpointing algorithm in Section 3 and different examples in Section 4. The correctness proof is provided in Section 5. In Section 6, we evaluate the proposed scheme. Section 7 presents conclusions.

## 2. RELATED WORK

Three classes of checkpointing protocols have been proposed for distributed systems: Coordinated, independent and communication induced. *Coordinated checkpointing* is a commonly used technique for fault tolerant [1-4,8,11,13,15,17,22,24-26]as it is domino free. In coordinated or synchronous checkpointing, processes must coordinate their checkpointing activities and take checkpoints in such a manner that the resulting global state is consistent. Therefore, coordinated checkpointing suffers from high coordination overhead associated with the checkpointing process. Mostly it follows two-phase commit structure. In the first phase, processes take tentative checkpoints and in the second phase, these are made permanent. The main advantage is that only one permanent checkpoint and at most one tentative checkpoint is required to be stored. In the case of a fault, processes rollback to last checkpointed state [7]. The Chandy-Lamport [6] algorithm is the earliest non-blocking all-process coordinated checkpointing algorithm. In this algorithm a marker are sent along all channels in the network and requires FIFO channels. In coordinated algorithm we may be require piggybacking of integer csn( checkpoint sequence number) on normal messages [1,2,8,19,22].

In *independent checkpointing*, processes do not synchronize their checkpointing activity and processes are allowed to records their local checkpoints in an independent way. After a failure, system will search a consistent global state by tracking the dependencies from the stable storage. The main advantage of this approach is that there is no need to exchange any control messages during checkpointing. But this requires each process to keep several checkpoints in stable storage and there is no certainty that a global consistent state can be built. It may require cascaded rollbacks that may lead to the initial state due to domino-effect [7]. Acharya and Badrinath[5] were the first who present a uncoordinated checkpointing algorithm for mobile computing systems. In their algorithm, an MH takes a local checkpoint whenever a message reception is preceded by a message sent at that MH. If the *send* and *receive* of messages are interleaved, the number of local checkpoints will be equal to half of the number of computation messages, which may degrade the system performance.

In *communication induced* checkpointing approach, a global checkpoint is similar to the approach of coordinated checkpointing while rollback propagation can be avoided by forcing additional un-coordinated local checkpoint in processes [11,12,26].

In [4,9,24]authors proposed blocking algorithms to minimizing the number of synchronization message and number of checkpoints during checkpointing. However, these processes force all relevant processes to block their underlying computation during the checkpointing process. Therefore, blocking algorithm may degrade the performance to mobile computing systems [8].

In [4] Koo and Tong's use a sequential coordinated scheme in which initiator node sends checkpoint request sequentially through a checkpoint dependency tree, from the root to the leaves and acknowledgement procedure is also sequential from the processes at the leaf levels through their ancestors, one level at a time, until the root (initiator) processes receives all processes.

Further to remove blocking overhead in [8,22] authors proposed all process non blocking centralized checkpointing algorithms with minimum synchronization message overhead. But these algorithms suffer form centralized algorithms disadvantages and require all process in the system to take checkpoint, even though many of them may not be necessary. When we modify these algorithms in distributed, these algorithms will suffer from another problem as mention in [3].

Recently, non-blocking distributed checkpointing algorithms [21,25] have received consideration attention. However, these algorithm [21,25] also forces all processes as in [4,9,24], even though many of them may not be necessary.

As mobile computing faces many new challenges such as low wireless bandwidth, frequent disconnections and lack of stable storage at mobile nodes. These issues make traditional checkpointing techniques unsuitable to checkpoint mobile distributed systems [1,5,15]. Minimum process Coordinated checkpointing is widely used technique in mobile distributed system as it requires less storage, bandwidth and have the characteristic of domino-free. To take a checkpoint, an MH has to transfer a large amount of checkpoint data to its local MSS over the wireless network. Since the wireless network has low bandwidth and MHs have low computation power, all-process checkpointing will waste the scarce resources of the mobile system on every checkpoint.

Hence, the problem of minimizing the number of synchronization messages and checkpoints is become a crucial issue in mobile system as wireless network has limited bandwidth and mobile nodes have limited computation, storage and energy conservation requirement. It is mostly desirable that a coordinated checkpoint algorithm forces a minimum number of processes to take checkpoints [14].

The Parkash-Singhal[15] proposed the first minimum process non blocking checkpointing algorithm. This algorithm only forces the minimum number of processes to take checkpoints without blocking of the underlying computation. However author found that this algorithm may result in an inconsistency [3,13] in some situation and proved that there does not exist a non-blocking algorithm which forces only a minimum number of processes to take their checkpoints.

Cao and Singhal [1] achieved non-intrusiveness in the minimum-process algorithm by introducing the concept of mutable checkpoints. If any process sends a computation message to another process after receiving the checkpoint request, the receiving process first take the mutable checkpoint first and process the message. Later, this mutable checkpoint converted to tentative if it receives checkpoint request related to the current initiation; otherwise it become the useless checkpoint. The number of useless checkpoints in [1] may be exceedingly high in some situations [19].

Kumar et. al [2] proposed a five phase checkpointing algorithm to reduced the height of the checkpointing tree and the number of useless checkpoints by keeping non-intrusiveness intact. It follows the following steps in a distributed system which has (n+1) processes. (i) Initiator process broadcasts the dependency vector request to all processes. (ii) Receives the dependency vector from all processes and then initiator process compute minimum set of processes which are directly or transitively dependent on the initiator process. (iii) Take own tentative checkpoint and send the tentative checkpoint request to the processes which belongs to the minimum set.(iv) Initiator process receives the responses of taking tentative checkpoint (v) initiator process send the commit or abort message to all the processes. However, this algorithm reduces the useless checkpoint in the comparison of algorithm [1] but has extra message overhead cost.

A good checkpoint algorithm for mobile systems needs to have following characteristics [10]: It should impose low memory overheads on MHs and low overheads on wireless channels. The disconnection of MHs should not lead to infinite wait state. The checkpointing algorithm should avoid awakening of an MH in doze mode operation. The algorithm should be non-intrusive and minimum-process. In minimum-process checkpointing algorithm, a process takes its permanent checkpoint only if the initiator process is directly or transitively dependent upon it.

## 3. THE PROPOSED CHECKPOINTING ALGORITHM

### 3.1 System Model

Our system model is similar to [1,19]. There are *n* spatially separated sequential processes denoted by $P_0, P_1,.., P_{n-1}$, running on MHs or MSSs, constituting a mobile distributed computing system. Each MH/MSS has one process running on it. The processes do not share memory or clock. Message passing is the only way for processes to communicate with each other. Each process progresses at its own speed and messages are exchanged through reliable channels, whose transmission delays are finite but arbitrary. A process is in the cell of MSS means the process is either running on the MSS or on an MH supported by it. It also includes the processes of MHs, which have been disconnected from the MSS but their checkpoint related information is still with this MSS. We also assume that the processes are non-deterministic.

### 3.2 Minimum Set and Maintenance of Dependency Vector

In order to maintain the dependency vector $ddv_i[]$, we use the similar approach as the [15], where each process $P_i$ maintains a Boolean vector $ddv_i[]$, which has n bits. Initially at $P_i$, the vector $ddv_i[]$set to 0 except $P_i[i]$ and set $ddv_i[j]$ to '1' only if $P_i$ receive computation message(m) from $P_j$. So, $ddv_i[j]$ =1 represents that $P_i$ is directly dependent upon $P_j$ for the current CI.

When process $P_i$ sends a computation message m to $P_j$, it appends $ddv_i[]$ to m. After receiving m, $P_j$ includes the dependences indicated in $ddv_i[]$ into its own $ddv_j[]$ as follows: $ddv_j[k]$ = $ddv_j[k]$ v m. ddv[k] , where 1<=k<=n, and v is the bitwise inclusive OR operator. Thus, if a sender $P_i$ of a message depends on a process $P_k$ before sending the computation message, the receiver $P_j$ also depends on $P_k$ through transitivity. So in this way ddv[] contain all the processes which are directly or transitively dependent on the process.

Minimum set is a bit vector of size n which is compute by the $MSS_{ini}$ by taking transitive closure of dependency of dependency bit vector with its own dependency bit vector. So at the time of initiation ddv [] of the $MSS_{ini}$ treated as a minimum set. (minset[]= $ddv_{ini}[]$). minset[k]=1 implies $P_k$ belongs to the minimum set and it is directly or transitively dependent on initiator process $P_{ini}$.

### 3.3. Data Structures

Here, we describe the data structures used in the proposed checkpointing protocol. A process on MH that initiates checkpointing, is called initiator process and its local MSS is called initiator MSS. If the initiator process is on an MSS, then the MSS is the initiator MSS. A process is in the cell of MSS means the process is either running on the MSS or on an MH supported by it. It also includes the processes of MHs, which have been disconnected from the MSS but their checkpoint related information is still with this MSS. All data structures are initialized on the completion of a checkpointing process if not mentioned explicitly.

*$P_{ini}$:* Initiator process identification.

*$MSS_{ini}$:* Initiator MSS identification.

*g_chkpt:* On the receipt of checkpoint initiation request $MSS_{ini}$ set flag to '1'. If it is already 1 it mean that some global checkpoint recording is already going on and in such case $MSS_{ini}$ discard the checkpoint initiation request.

*$weight_i$:* A non negative real variable with a maximum value of 1and used to detect the termination of checkpointing algorithm as in [10]. $MSS_{ini}$ attach some portion of the weight along with checkpoint request when it sends c_req to all process which belongs to minset[].

*$mr_i$:* When any process $P_i$ receive the checkpoint request message(c_req) it will reply positively or negatively. So a flag set to "1" on taking the tentative or induced checkpoint successfully.

*$csn_i[]$:* An array of length n for n processes at each process $P_i$, where $csn_i[j]$ indicates the checkpoint sequence numbers (csn) of $P_j$ currently known to $P_i$.

*$own_csn_i$:* The csn of $P_i$'s last checkpoint. The csn of process $P_i$ increases monotonically. So on switching chk_state own_csn= csn[i] +1 and on commit or abort $own_csn_i$=csn[i].

*$ddv_i[]$:* A bit vector of size n; $ddv_i[j]$ =1 implies $P_i$ is directly dependent upon $P_j$ for the current CI; $ddv_i[j]$ is set to '1' only if $P_i$ processes m received from $P_j$ s.t. m.own_csn □ $csn_i[j]$; otherwise $ddv_i[j]$=0. m.own_csn is the own_csn at $P_j$ at the time of sending m and $csn_i[j]$ is $P_j$'s recent permanent checkpoint's csn; initially, k, $ddv_i[k]$=0 and $ddv_i[i]$=1; for $MH_i$ it is kept at local MSS; maintenance of ddv[] is described in sections 2.2

*c_state:* A flag set to "1" when a process $P_i$ takes tentative after receiving the ckreq_msg or some condition mentioned in 2.4.c

*$Sendv_i[]$:* A bit vector size n; $sendv_i[j]$=1 implies $P_i$ has sent at least one message to $P_j$ in the current CI.

*minset[]:* A bit vector of size n which is compute on the $MSS_{ini}$ ; if Pi initiate its(x+1)th checkpoint then the set of processes on which Pi depends(directly or transitively) in its $x^{th}$ checkpoint interval is minimum set. $MSS_{ini}$ computes minset[](subset of minimum set) on the basic of ddv[] maintained at $MSS_{ini}$. initially minset[]= $ddv_{ini}[]$. So minset[k]=1 implies $P_k$ belongs to the minimum set and it is directly or transitively dependent on initiator process $P_{ini}$ . In order to compute the initial minimum set we use the similar approach as [15].

*$new_ddv_i[]$* it hold the new dependency at node $P_i$ during the execution of checkpoint request.

*Uminset[]:* On receiving $new_ddv_k[]$ from some $MSS_k$ with response, Uminset[] is updated by (minset[])U($new_ddv_k$[]). It contains the exact minimum set; 'U' is a operator for bitwise logical OR; new_ddv[] is describe above.

*c_req:* Initiator and other node send Checkpoint request to their dependent

*c_rply:* After receiving the checkpoint request, processes send reply(acknowledge) negatively or positively to the initiator directly.

## 3.4 Minimum Process Coordinated Checkpointing Algorithm

Each process $P_i$ can initiate the checkpointing process. If $MH_i$ initiates checkpointing, it sends the request to its current MSS (initiator MSS) that initiates and coordinates checkpointing process on behalf of $MH_i$. If some checkpointing activity is already going on (*g_chkpt* is set at the initiator MSS), then the new initiation is ignored.

When an MH sends an application message, it is first sent to its local MSS over the wireless cell. The MSS piggybacks appropriate information with the application message, and then routes it to the destination MSS or MH. Conversely, when the MSS receives an application message to be forwarded to a local MH, it first updates the data structures that it maintains for the MH, strips all the piggybacked information, and then forwards the message to the MH. Thus, an MH sends and receives application messages that do not contain any additional information; it is only responsible for checkpointing its local state appropriately and transferring it to the local MSS.

### 3.4.1 On Checkpoint Initiation:

Each process can initiate a checkpoint. When $P_i$ initiates a checkpointing algorithm it follow (i)set the global checkpoint initiation to '1' (ii) sets tentative local checkpoint to '1' (iii) increment in own checkpoint sequence number(csn) (iv) sets weight to '1'(v) add own process , MSSs identifier and current interval number in Master_ set.(vi) compute minset[] of processes which are directly or transitively dependent on initiator. (vii) At last, sends the checkpoint request(c_req) to all the member of minset[] with Master_set, minset, ws and request.

### 3.4.2 On reception of checkpoint request:

On the receipt of checkpoint request it depends upon the processes whether he willing to take tentative checkpoint or not. If process willing to take checkpoint set, it set mr= =1 else set '0'. When $P_j$ receive message c_req() from $P_i$ , it will first compare the $csn_j$ and Master_set with the req.own_csn and req.Master respectively. If $csn_j$ >req.own_csn or $own_Master_j$ =req_Master, then $P_j$ does not take a checkpoint and if not, then $P_j$ takes tentative checkpoint. Here, $Master_j$=req_Master means received request is duplicate and $P_j$ already taken a tentative checkpoint related to the same initiator. $P_j$ check its own direct dependency vector($ddv_j$) and any process may exist in one of the following case:

Case 1: ($ddv_j$[] = $\varnothing$ ) ;
Case 2: ($ddv_j$[] = = minset[] ;
Case 3: ($ddv_j$[] $\neq$ minset[]) $\wedge$ (mr = = 1)

a) for (some k s.t. $ddv_j$[k] = = 1 $\wedge$ minset[k]= = 1)
b) for (some k s.t. $ddv_j$[k] = = 1 $\wedge$ minset[k]= = 0)
   i. $sendv_j$[] = = $\varnothing$ ;
   ii. some p s.t. $sendv_j$[k] = = 1 $\wedge$ minset[k]= = 0;
   iii. some p s.t. $sendv_j$[k] = = 1 $\wedge$ minset[k]= = 1;

After taking the tentative checkpoint, $P_j$ needs to send checkpoint request to those processes which are directly or transitively dependent on it. $P_j$ finds out those dependent processes which are not part of minimum set and not sent any message to the processes which are belongs to minimum set. If the above condition true, it set $new_ddv_j$=1 and sends the checkpoint request to such processes with some portion of weight and sends reply to initiator MSS with remaining weight. If not, sends reply to the initiator MSS with received weight.

if(case1) $\vee$ (case2) $\vee$ (case 3a) $\vee$ (case 3b. iii)
  {Sends message c_rply() with weight received and message response.}
else if ((case 3b. i) $\vee$ (case 3b. ii)) $\wedge$ mr = =1
  { Set $new_ddv_j$[k]= = 1;
    wr = wr/2; ws = wr ;
    $P_j$ sends message c_req(trigger_set, minset[], ws, REQUEST) to process $P_k$
    $P_j$ sends message c_rply($new_ddv_j$[], wp, wr) to $MSS_{ini}$ }

### 3.4.3 Computation message received during checkpointing:

During checkpointing when a process $P_i$ receives computation message from $P_j$ , to main maintain the consistent global state $P_i$ either (i) takes tentative checkpoint before processing the message or (ii) takes tentative checkpoint after processing the message or (iii) $P_j$ buffer the message. When any process sends computation message after taking the tentative checkpoint (i.e. m.state = =1), it attach minset[], m.own_csn$_j$ and m.state$_j$ = = 1 with the message .
After receiving the message from $P_j$, $P_i$ compares m.own_csn with its local csn$_i$[j]. if m.own_csn <= csn$_i$[j], (means message sent without taking checkpoint)the message is received without taking any checkpoint. Otherwise, $P_i$ takes checkpoint and receives the message so that the message not becomes orphan. $P_i$ also update its csni[j] and sets c_state$_i$ '0' to '1'.

The following steps are happen as per the given conditions(as given in 2.5.5):

### 3.4.4 Termination of Checkpointing Algorithm

When a $MSS_{ini}$ receives newddv[] and weight with message c_rply(), it adds the newddv[] in to new_set[] and weight received in to its own weight. Uminset is calculated by taking the union of minset[] and newest[]. $MSS_{ini}$ sends message ABORT() if one of the following condition occur (i) mr= = 0, if at least one of its relevant process has failed / not willing to take its tentative checkpoint and sends message response negatively (ii) Time out, if $MSS_{ini}$ not receive responses from all processes within the given time. Initiator MSS commits only if every relevant process/processes take its tentative checkpoint. When its weight becomes equal to 1 as in [10], initiator MSS say $MSS_{ini}$ concludes that all of its relevant processes have taken the tentative local checkpoint successfully. Finally, initiator MSS sends COMMIT/ABORT to all processes belongs to Uminset[].On receiving ABORT: processes discard the tentative checkpoint and on receiving COMMIT: processes convert its tentative checkpoint into permanent one and discard its earlier permanent checkpoint, if any; otherwise, it processes the buffered messages. When any MSS which belongs to Uminset[] receive the COMMIT or ABORT from the initiator MSS then it sends the request to all its processes related to the current checkpoint initiation and update its own data structure.

## 3.5 Formal Outline of the minimum-process Algorithm:

### 3..5.1 Algorithm executed on an initiator (Pi)

a) if *$P_{ini}$ runs on an MH*
  {$P_{ini}$ sends checkpoint initiation request to its local MSS say $MSS_{ini}$}
b) if*( g_ckpt= =1)* // some global checkpoint recording is already going on
  {Discard checkpoint initiation request; Inform to initiator process; Exit ;}
c) *on checkpoint initiation:*
  set g_ckpt$_i$ = 1; new_set[]= = 0; Uminset[]= = minset[];
  set c_state$_i$=1; // take tentative local checkpoint
  own_csn$_i$= own_csn +1; // increment in own_csn
  set weight$_i$=1.0;
  check ddv$_i$ ; when ddv$_{ini}$[k] = =1 for 1<=k<=n; set minset[k]=1;
  set trigger($P_{ini}$, $MSS_{ini}$, own_csn$_i$,);
d) *sends c_req() to all node $P_j$ such that*
  for(j=0;j<=n; j++)
    { if (minset[j]= = 1)
    weight$_i$ = weight$_i$/2; ws=weight$_i$ ;
    sends c_req(trigger, minset,ws,REQUEST);
    }
e) continue normal operation. If any checkpoint initiation request is received discard
  it and continue normal operation;
f) *on receiving response from process Pj:*
  Receive message c_rply (new_ddv$_j$[],w$_j$, mr)

```
if(new_ddvj[] ≠ ∅ )
      {new_set = new_set[] U new_ddvj[];
       Uminset = minset[] U new_set[];}
if(weight<1) ∨ (maxtimeout)
      {if mr= =0 // check mr
            send abort() to all processes belongs to Uminset[]
       else
            weighti = weighti + wj ;}
if(weight= =1)
      { Send message COMMIT() to all process belongs to Uminset[];}
```

### 3.5.2 When any process Pj is element of minset AND receive(c_req)

```
a)  if req_trigger.Pid = own_trigger.Pid // Pj has already taken checkpoint related to CI
        { ignore the checkpoint request}
b)  if req.trigger ≠ own_trigger // Pj has not taken checkpoint related to CI
        { Take tentative checkpoint ; increment csnj ; check ddvj[];
       i)  if(ddvj[] = = ∅ ) ∨ (ddvj[] = = minset[])
          {wr = ws;
           Sends message c_rply(∅, wr, mr) to MSSini
           Continue computation ;}
       ii) if(ddvj[] ≠ minset[]) // sends c_req only if new dependency occure
           a)  if(∃k s.t. ddvj[k] = =1 ∧ minset[k] = =0) // new dependency
                  i) if(sendvk[] = =∅) ∨ (∃p s.t. sendvk[p] = =1 ∧ minset[p] = =0)
                      { wr = wr/2; ws=wr;
                        Pj sends message c_req(minset[],csnj, req.trigger,ws
                                                        REQUEST) to process Pk ;
                        Pj sends message c_rply(new_ddvj[k],wr, mr) to MSSini;
                        Continue computation;}
                  ii) if(∃p s.t. sendvk[p] = =1 ∧ minset[p] = =1) // No new dependency
                        { No c_req() are sent further; continue computation;}
           b)  if (∃k s.t. ddvj[k] = =1 ∧ minset[k] = =1) // No new dependency
                  {No need to send c_req(); continue computation;}
```

### 3.5.3. When any process $P_j$ is about to send out a message to $P_i$

```
Set sendvj[i] =1;
if(m.c_statej = = 0)
{Send(Pi, message, ddvj[],own_csnj, null);}
else
{send(Pi, message, ddvj[], own_csnj, minset[]);}
```

### 3.5.4 When Pi receive a computation message from Pj

```
a) if(own_csn <= csni[j]) ∨ (m.statej = = c.statei)
       { Receive(m) and update csni[j] and ddvi[j];} /* nobody need to take checkpoint
         as both processes have taken or not taken checkpoint related to the CI */
b) if(own_csn <= csni[j]) ∨ (m.statej = = 1 ∧ c.statei = = 0)
     i) if(Pi ∈ minset[]) ∧ (c_state= = 0) // not receives c_req() yet
          {Take tentative checkpoint; receiving message;
            increment own_csni; set trigger; update csni[j];update ddvi[j]; }
     ii) if(Pi ∈ minset[]) ∧ (c_state= = 1) // Pj already participated in chkpt algo
          { Receive(m) and update csni[j] and ddvi[j]; }
```

iii) if (($P_i \notin$ minset[]) $\wedge$ (Bitwise logical AND of $sendv_i$[] $\wedge$ minset[] is all zero))
{Buffer the message}// this buffered message execute after chkpt session
iv) if (($P_i \notin$ minset[]) $\wedge$ (Bitwise logical AND of $sendv_i$[] and minset[] is not all zero) // there is a good probability that it will get c_req() in future
{Take tentative checkpoint ; receiving message; increment $own_csn_i$; set trigger, update $csn_i[j]$; update $ddv_i[j]$ }
v) if ($P_i \notin$ minset[]) $\wedge$ ($sendv_i$[] = = $\varnothing$) // not sent any message
{receive message; }

### 3.5.5 When any MSS which belongs toUminset[] receive COMMIT or ABORT :

a) send the request COMMIT/ABORT to all of its local processes;
b) update data structures;

### 3.5.6 Algorithm Executed at Any Process Pi:

a) *Upon receiving a tentative checkpoint request from local MSS:*
   i) Takes tentative checkpoint. If already taken tentative checkpoint ignore the request.
   ii) Send the response positively or negatively to its local MSS;
b) *On receiving Commit( )*
   **if** (*tentative*$_i$) {
   {discard old permanent checkpoint, if any;
   convert the tentative checkpoint into permanent one;}
c) *On receiving Abort ( )*
   **if** (*tentative*$_i$)
   {discard the tentative checkpoint;}

## 3.6 Handling node mobility, disconnections and failure during checkpointing

Due to mobility a MH may disconnect from the old MSS and connected to a new MSS. Due to this message transmission becomes complicated. Many routing protocol have been proposed [16,18] to handle the MH mobility.

Disconnection of an MH is a voluntary operation [5], and frequent disconnections of MHs are an expected feature of a mobile distributed system. An MH may be disconnected from the network for an arbitrary period of time. The Checkpointing algorithm may generate a request for such MH to take a checkpoint. Delaying a response may significantly increase the completion time of the checkpointing algorithm. We propose the following solution to deal with disconnections that may lead to infinite wait state.

Suppose, an MH, say $MH_i$, disconnects from the MSS, say $MSS_k$. $MH_i$ takes its checkpoint, say $disconnect_ckpt_i$, and transfers it to $MSS_k$. $MSS_k$ stores all the relevant data structures and $disconnect_ckpt_i$ of $MH_i$ on stable storage. If $MH_i$ is in the minset[], $disconnect_ckpt_i$ is considered as $MH_i$'s checkpoint for the current initiation. On commit, $MSS_k$ also updates $MH_i$'s data structures, e.g., ddv[], send, etc. On the receipt of messages for $MH_i$, $MSS_k$ does not update $MH_i$'s *ddv*[], but maintains a message queue to store the messages.

When $MH_i$ enters in the cell of $MSS_j$, it is connected to the $MSS_j$ if no checkpointing process is going on. Before connection, $MSS_j$ collects its ddv[], buffered messages, etc. from $MSS_k$; and $MSS_k$ discards $MH_i$'s support information and $disconnect_ckpt_i$. The stored messages are processed by $MH_i$, in the order of their receipt at the MSS. $MH_i$'s ddv[] is updated on the processing of buffered messages. If a node does not reconnect in a stipulated time, then its computation can be restarted from its disconnect_ckpt. There is also a possibility that, during checkpointing activity, an MH fails and all processes running on it also fail. In our proposed algorithm failure are handled as in[1].

## 4. EXAMPLES

### Example 1. (Sending and receiving computation message)

We explain our checkpointing algorithm with the help of an example. Consider the distributed system as shown in Fig. 2. This mainly shows the receiving of the computation message during checkpointing. Note that when a computation message is sent after taking the checkpoint it piggybacked with minset[]. Assume that process $P_2$ initiate checkpointing process. First process $P_2$ takes its tentative checkpoint $C_{2,1}$ and updates own_csn$_2$ to 2, compute minset[][which in case of Fig 2. is {$P_0$,$P_1$, $P_2$, $P_3$}]. This means that process $P_0$ , $P_1$ and $P_3$ sends at least one message to process $P_2$ and since $P_2$ has already taken its checkpoint $C_{2,1}$ these message become orphan if $P_0$, $P_1$ and $P_3$ do not take checkpoint. Hence, when $P_2$ initiate a checkpoint all of these processes which are directly or transitively dependent on $P_2$ should take their checkpoints in order to maintain global checkpoint consistency. Therefore P2 sends the checkpoint request along with minset[] to process $P_0$, $P_1$ and $P_3$.

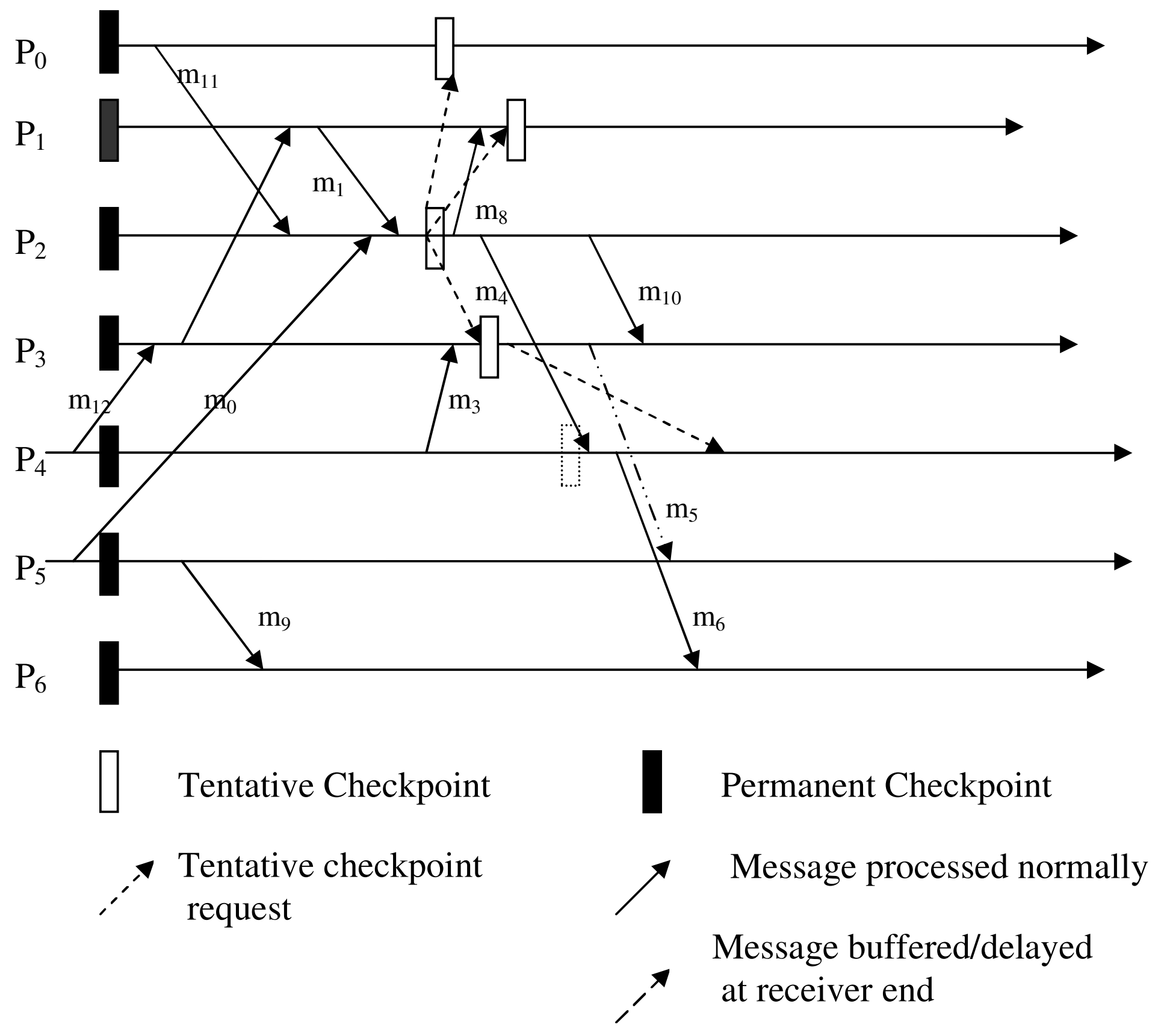

**Fig. 2.** Sending and receiving computation message during checkpointing

When $P_0$ receives the checkpoint request it takes the tentative checkpoint. $P_2$ sends $m_8$ with minset [] after taking its checkpoint (m.c_state$_2$ = =1) and $P_1$ receives $m_8$ before getting the minset[]. In this case, $P_1$ takes tentative checkpoint and receives $m_8$ (*IVb.i*) as minset[$P_1$]= =1. so, it knows that it is the part of minset and get the checkpoint request from the $MSS_{ini}$ and when it get the checkpoint request it ignore the request as c_state$_1$ = = 1. After taking its checkpoint, $P_2$ sends $m_4$ to $P_4$. As minset[$P_4$]= = 0 ,means $P_4$ does not belongs to minset, $P_4$ takes bitwise logical AND of sendv$_4$[] and minset[] (*IVb.iv*)and finds that the resultant vector is not all zeroes [sendv$_4$[3]=1 due to $m_3$; minset[3]=1]. $P_4$ concludes that most probably, it will get the checkpoint request in the current initiation; therefore, it takes its tentative checkpoint before processing $m_4$.When $P_3$ takes its tentative checkpoint, it finds that it is dependent upon $P_4$ and $P_4$ is not in the minimum set [known locally]; therefore, $P_3$ sends checkpoint request to $P_4$ and send reply to the $MSS_{ini}$ with new_ddv$_3$ [4] =1. $MSS_{ini}$ compute the Uminset[$P_0$, $P_1$, $P_2$, $P_3$, $P_4$] by taking the union of minset{$P_0$, $P_1$, $P_2$, $P_3$} and new_ddv$_3$[4]. Later when $P_4$ receive the checkpoint request from $P_3$ it ignore the request and continue the normal operations.

After taking its checkpoint, $P_3$ sends $m_5$ to $P_5$. $P_5$ takes the bitwise logical AND of sendv$_5$ [] and minset[] (*IVb. iii*) and finds the resultant vector to be all zeroes (sendv$_5$[]=[000001]; minset[]=[111000]). $P_5$ concludes that most probably, it will not get the checkpoint request in the current initiation; therefore, $P_5$ does not take tentative checkpoint but buffers $m_5$. $P_5$ processes $m_5$ only after getting commit request. $P_6$ processes $m_6$ *(IVb.v)*, because, it has not sent any message since last permanent checkpoint. After taking its checkpoint, $P_2$ sends $m_{10}$ *(IVb.ii)* to $P_3$. $P_3$ processes $m_{10}$, because, it has already taken its checkpoint related to the current initiation.

At last, when $P_2$ receives positive responses from all relevant processes(weight = =1) it issues commit request along with the exact minimum set [$P_0$, $P_1$, $P_2$, $P_3$, $P_4$ ] to all processes. On receiving commit following actions are taken. A process, in the minimum set, converts its tentative checkpoint into permanent one and discards its earlier permanent checkpoint, if any. Processes the buffered messages, if any. csn[], ddv[] and other data structures are updated. Hence, a process takes tentative checkpoint only if there is a good probability that it will get the checkpoint request in the current initiation; otherwise, it buffers the received messages. In this way, our proposed algorithm tries to optimize the number of resources without taking useless checkpoints and blocking of processes. On the other hand if $MSS_{ini}$ receive the negative response from any one of the processes which belongs to the minset, it sends the abort message to all processes which belongs to Uminset[]. On receiving abort, processes discard the tentative checkpoint, if any; reset c_state, tentative, g_chkpt etc and update ddv[] and minset[].

**Example 2. (Sending and receiving checkpointing request during checkpointing)**

In this example we show that how the checkpoint request are sent and receives during checkpointing in our proposed checkpointing algorithms. Consider another distributed system as shown in the Fig. 3. Here $P_2$ initiate the checkpointing algorithm. First process $P_2$ takes its tentative checkpoint, increment its csn$_2$ from 1 to 2 and check its minset[] which is { $P_1$, $P_2$, $P_3$, $P_4$}. So, to maintain the consistent global state $P_2$ sends tentative checkpoint request along with minset[$P_1$, $P_2$, $P_3$, $P_4$] and csn$_2$= 1 and trigger set to all processes which belongs to minset.

When $P_1$ receives the checkpoint request from the $MSS_{ini}$ , if first compare the received trigger $P_{id}$ with its own trigger $P_{id}$ and find that these are not equal. So $P_1$ takes checkpoint , increment csn$_1$ and set trigger equal to request trigger, then checks ddv$_1$[] which is null. Hence $P_1$ sends reply (not shown in figure) to the $MSS_{ini}$ with weight received *(IIb.i)* and continue normal operation.

Similarly processes $P_3$ and $P_4$ first take tentative checkpoint and increments its csn$_3$ and csn$_4$ respectively, then each process checks its dependency vector to find the dependent process.Here ddv$_3$[] is {$P_4$,$P_5$,$P_6$} and $P_3$ checks all the processes one by one. Taking $P_4$, $P_3$ finds that it belongs to minset. So $P_3$ does not sends further checkpoint request to $P_4$ as it receives the checkpoint request directly from the $MSS_{ini}$ *(IIb.iib)*. Now taking $P_5$, as $P_5$ does not belongs to minset but it (sendv$_5$[$P_4$]=1 and minset[$P_4$] =1)has sent a message to process $P_4$ which belongs to minset[](IIb.iia.ii). Hence $P_3$ does not send further checkpoint request to process $P_5$ and continue normal operations, as it will get/gotten the checkpoint request from the process $P_4$. At last, as $P_6$ not belongs to minset and it has not sent any message ($sendv_6[P_7]$=1 and minset[$P_7$]= 0) to the process which belongs to minset*(IIb.iib)***.** So, $P_3$ sends the checkpoint request with some portion of the weight to $P_6$. it also sends reply(not shown in figure) to the $MSS_{ini}$ with message response, remaining weight and new_ddv$_3$[$P_6$].

When $P_6$ receives the checkpoint request it takes checkpoint and send the reply(not shown in figure) directly to the $MSS_{ini}$. When $MSS_{ini}$ receaves reply from the processes it adds the weight in its own weight and new ddv[] in its own new_set[]. Same in case of $P_4$ when it receive checkpoint request and sends checkpoint request to $P_5$ as $P_3$ receives checkpoint request and sends checkpoint request to $P_6$. By this way new_set[]= new_set[] U new_ddv[]; Uminset[]=minset[] U new_set[] at last $P_2$ gets the exact Uminset[$P_1$, $P_2$, $P_3$, $P_4$, $P_5$, $P_6$] . When weight become1, means $P_2$ receives responses from all its v relevant processes and $P_2$ issues commit along to all processes which belongs to Uminset[]. If any message is sends after taking tentative checkpoint, it will be handled as fig. 2.

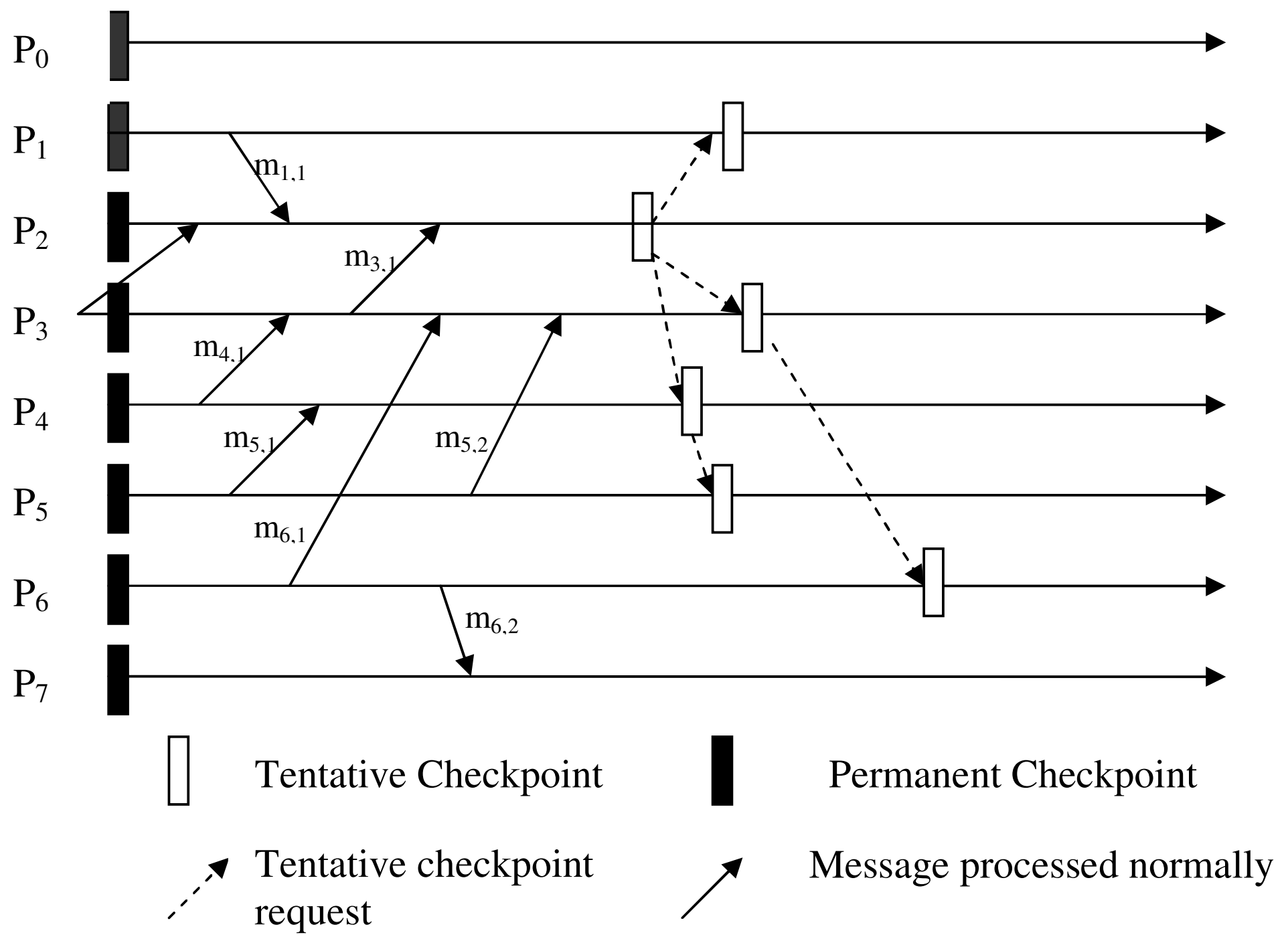

Fig. 3. Sending and receiving checkpointing request during checkpointing

**Example 3. Handling of Tardy Messages**

Our proposed algorithm tries to maintain the dependency information available right at the time when a process initiates a checkpoint as section 2.5. So we send the computation message by attaching the checkpointing-dependency information of the sending process to the destination process. It is important to note that during normal message transmission, attached $ddv_j$[] contains the processes from which it receive at least one computation message before sending message to the destination process. So $ddv_j$[] hold accurate dependency it there is no tardy message in the system. A message is tardy if it is received by a process P after P has send out at least one message [14]. This example shows the problem and solution of tardy message as per our proposed algorithms.

As per Fig. 4. $P_1$ first sends a message $m_{1,1}$ to $P_2$, then $P_2$ sends a message $m_{2,1}$ to $P_3$, finally $P_0$ sends a message $m_{0,1}$ to $P_1$. So $P_1$sends $ddv_1$ is equal to null with computation message $m_{1,1}$ to process $P_2$, as it does not receives any computation message yet. $P_2$ sends computation message $m_{2,1}$ to $P_3$ by attaching $ddv_2$ is equal to $\{P_2\}$ as it receives the message $m_{1,1}$ from process $P_1$.Hence, when $P_3$ initiates a checkpoint, all of three processes should take their checkpoint in order to maintain the consistency. However, when $P_3$ initiate a checkpoint, $ddv_3$ is equal to $\{P_1,P_2\}$, rather than $\{ P_0, P_1,P_2\}$.

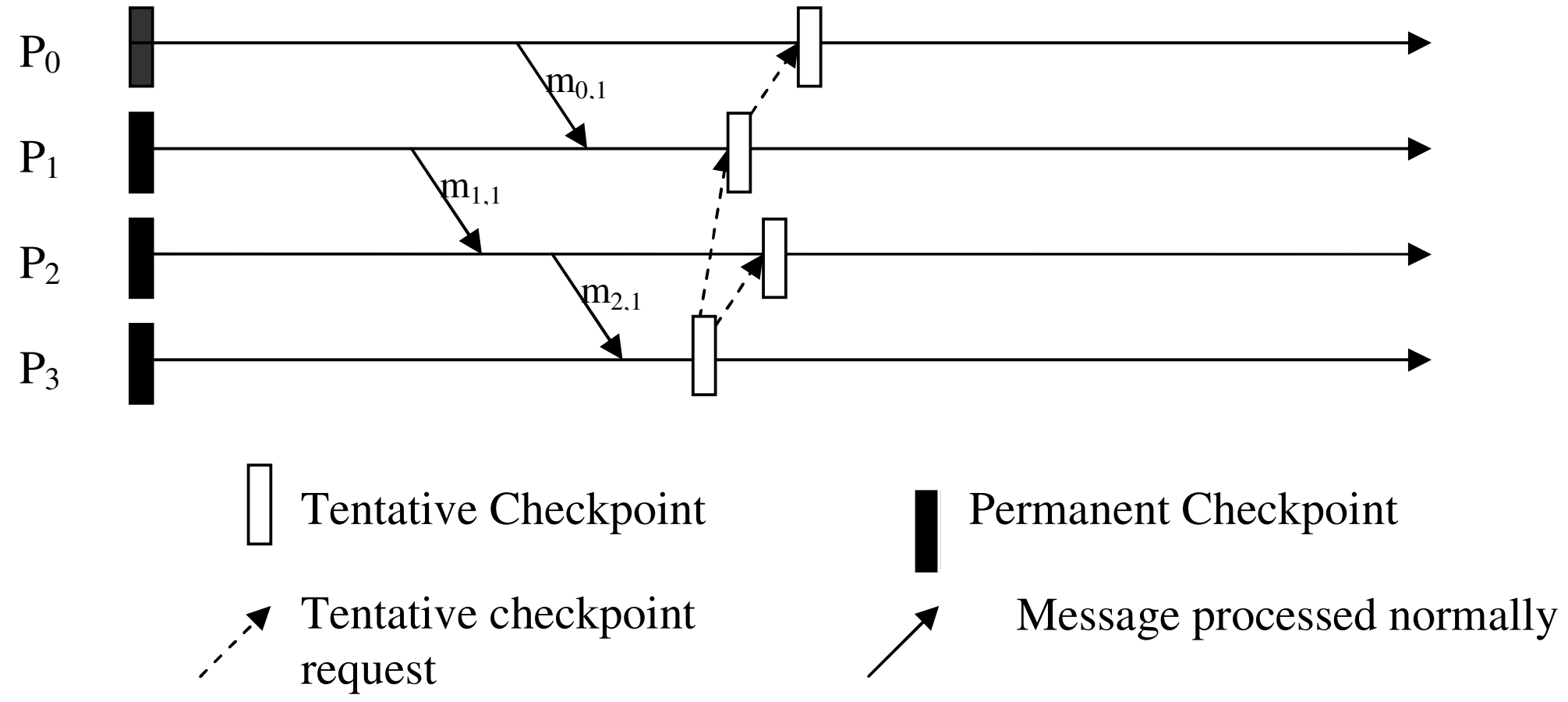

Fig. 4. An example showing tardy message

This is because the fact that P has receives a message from $P_1$ after sending the message to process $P_3$. So $m_{0,1}$ is recorded in $ddv_1$, rather than $ddv_3$ by the time $P_3$ initiates a checkpointing. In other word, the message $m_{0,1}$ in Fig. 4. is a tardy message. This example shows that we need to handle such tardy message properly else it become orphan and produce inconsistent global state.

Our proposed checkpointing algorithm accurately identify all processes potentially involved in a global checkpoint without blocking of processes and work efficiently even in the presence of tardy messages. As per our proposed algorithm $P_4$ initiate the checkpoint algorithm and minset is equal to $\{ P_1,P_2,P_3\}$. Hence, $P_3$ takes tentative checkpoint and sends checkpoint request along with minset[] to processes $P_1$.

## 5. Correctness Proof

**Theorem I**: A process can not be a member of any minimum set, if it has not sent a message in its current checkpointing interval and minimum numbers of processes take checkpoint.

**Proof:** if $P_i$ is initiator and initiates its checkpointing algorithms then it contains all the processes in minset which are directly or transitively dependent on $P_i$ in current checkpoint initiation. So a process can not become the member of minset if it has not sent a message in current checkpoint initiation. Process $P_j$ will take a checkpoint related to current initiation if and only if it is directly or indirectly dependent or sends a computation message to a process which is directly or indirectly dependent on initiator. If dependency is counted at the time of initiation it will get the checkpoint request directly from the initiator and in case of any tardy message it will get the checkpoint request from that particular process in which it is directly dependent.

On the basic of theorem 1 we conclude that:

a) Each process notified by the global initiator or any process which are directly or indirectly after taking the checkpoint at most one checkpoint.
b) Our algorithms only forces minimal number of processes to take checkpoint.
c) If the set of checkpoint the checkpoint state is consistent before execution of our proposed algorithm, then it also consistent after the termination of algorithm.

**Theorem II:** Checkpoint algorithm terminates in certain period.

**Proof:** When initiator initiates a new checkpointing algorithm, the initiator and other processes follows the steps as (see section 2.5) and handled the disconnections and node mobility as (see section 2.6) and termination as (see section 2.4d).So all the nodes which are the part of global state must complete above steps in finite time unless a node is faulty or sends reply negatively to the $MSS_{ini}$. If a process in the minset become faulty or sends checkpoint message reply negatively during checkpointing, the whole of the checkpointing process is aborted. Hence, it can be inferred that the algorithm terminates in certain period.

**Theorem III:** Algorithm is non-blocking and produces a consistent global state

**Proof:** Processes which are part of global state can receives and handle the computation messages by the following ways

a) *Message received from the processes before sending any ddv[] with message to the initiator or any other processes which are directly or transitively depends upon initiator:* These types of processes become the part of the minset and receives the checkpoint request directly from the initiator so that these messages not become orphan.

b) *Message received from the processes after sending the ddv[] with message to initiator the initiator or any other processes which are directly or transitively depends upon initiator but before receiving the checkpoint request and taking tentative checkpoint.* Such types of messages are called tardy message and handled as (see tardy message).

c) Message received after taking tentative checkpoint and before receiving the commit request: These types of message are buffered and execute after the checkpoint interval.

In such way there are not any orphan message and handle all messages efficiently without blocking. So it shows that our algorithm is non-blocking and produces the global consistent state.

**Claim1:** Avalanche effect does not occur in our algorithm**:**

**Proof:** In case of avalanche effect chain of request form a loop as $P_i, P_j, P_k, \ldots P_s$, $P_i, P_j, P_k$. This chain does not occur in our algorithm. Suppose $P_i$ is initiator, it sends the checkpoint request along with minset to those processes which belongs to the minset. Let $P_j$ is the member of minset, so it get the checkpoint request from the initiator and take checkpoint. After taking checkpoint $P_j$ check its dependency, and sends checkpoint request further to $P_k$ on the behalf of initiator if and only if the following both the condition true in case of $P_k$:

a) $ddv_j[k] == 1 \wedge minset[k] == 0)$

b) $(sendv_k[] == \varnothing) \vee (sendv_k[p] == 1 \wedge minset[p] == 0)$

So there is not any possibility through which process which belongs to minset including initiator gets the checkpoint request again. Hence avalanche effect is not possible in our proposed algorithm.

## 6. Performance Evaluation of Proposed Checkpointing Algorithm

To evaluate we compare the performance of our proposed minimum process checkpointing algorithm with [1,2,4,8,25]in different perspective. We assume an n+1 process distributed system and use the following notations for performance analysis of the algorithms:

| | |
|---|---|
| *N:* | Total number of processes. |
| $N_{min}$: | Minimum number of processes that required to take checkpoint. |
| $N_{mut}$: | Number of redundant Mutable checkpoint during a checkpointing process. |
| $N_{indu}$ | Number of redundant Induced checkpoint during a checkpointing process. |
| $C_{broad}$: | Cost of broadcasting a message to all (N) processes in the system. |
| $C_{air}$ : | Cost of sending a message from one process to another process. |
| $T_{ch}$: | The checkpointing time. This time includes the time to save the checkpoint on MSS, transferring time from MH to its MSS and times taken by a system message during a checkpointing process. |

## 6.1. Performance of our proposed algorithm

**The blocking time**: Similar to algorithms [1,2,8,25], our algorithm does not block their underlying computation during checkpointing.
**The number of Checkpoints:** Similar to algorithms [1,2,4], our algorithm also forces only a minimum number of processes to take their checkpoints.
**The average message overhead:** our algorithm includes the following message overhead in best case is given as $3*N_{min} * C_{air}$ in table 1. In our algorithm, first the initiator sends control messages to minimum number of processes that need to take a checkpoint each and reply(acknowledge) back. At last when initiator receives the acknowledge from all the processes, it sends commit message to these minimum processes to convert their respective tentative checkpoint in to permanent one. Hence total cost of these are $3*N_{min}*C_{air}$.
**Useless Checkpoints:** Our algorithm does not have any useless checkpoint as [1][2].
Instead of above, our algorithm is coordinated, nondeterministic, distributed and require piggybacking of integer csn( checkpoint sequence number) on normal messages .

## 6.2. Comparison with Existing Algorithms

In [1], Cao-Singhal proposed a mutable checkpoint based non-blocking minimum-process coordinated checkpointing algorithm. This algorithm completes its processing in the following three steps. First initiator MSS sends tentative checkpoint request to minimum number of processes that need to take checkpoint. The synchronization message overhead for this is $N_{min} *C_{air}$. Secondly $MSS_{ini}$ gets the acknowledgement from all processes to whom it sent checkpoint request. Hence message overhead $2* N_{min} *C_{air}$ is needed in first two phases. At last $MSS_{ini}$ sends the commit request to convert its tentative checkpoint into permanent. In this case it takes $min(N_{min}* C_{st}, C_{broad})$. Hence algorithm [1] generate consistent global state with the message overhead cost $2* N_{min} * C_{air} + min(N_{min}* C_{air}, C_{broad})$ and average number of checkpoints $N_{min}+ N_{mut}$ [Refer Table 1]. Thus algorithm is non-blocking and minimum process but suffer from useless checkpoints. Our proposed algorithm generates the consistent global state with approximately same message overhead as [1], without using any useless checkpoint.

Table 1
A Comparison of System Performance opf the Proposed Minimum Process Check Pointing Algorithm

| | Cao-Singhal "Mutable"[1] | P.Kumar "Non-intrusive" [2] | Koo-Tong [4] | Elnozahy[8] | S.Neogy "CCUML"[25] | Our Algorithm |
|---|---|---|---|---|---|---|
| Average message overhead | $2* N_{min} * C_{air} + min(N_{min}* C_{air}, C_{broad})$ | $3*C_{broad} + 2*N_{min} * C_{air}$ | $3*N_{min}*N_{dep} * C_{air}$ | $2*C_{broad} + N*C_{air}$ | $2*C_{broad} + N*C_{air}$ | $3*N_{min} * C_{air}$ |
| Average no. of checkpoints | $N_{min}+N_{mut}$ | $N_{min}+N_{indu}$ | $N_{min}$ | N | $N$ | $N_{min}$ |
| Average blocking time | 0 | 0 | $N_{min}*T_{ch}$ | 0 | 0 | 0 |
| Useless checkpoint | Present | Present | No | No | No | No |
| Minimum Process | Yes | Yes | Yes | No | No | Yes |
| Non-blocking | Yes | Yes | No | Yes | Yes | Yes |
| Coordinated | Yes | Yes | Yes | Yes | Yes | Yes |
| Piggyback information | Integer | Integer | Integer | Integer | Integer | Integer |
| Concurrent executions | Yes | No | No | No | No | No |
| Distributed | Yes | Yes | Yes | No | Yes | Yes |

In [2], P.Kumar et al. also proposed minimum process coordinated checkpoint algorithm for mobile system. The synchronization message overhead to complete the checkpointing process using algorithm [2] is given as $3*C_{broad} + 2*N_{min} * C_{air}$. Here $3C_{broad}$ is the total cost of broadcasting sends ddv[]($C_{broad}$), take tentative checkpoint request($C_{broad}$) and commit($C_{broad}$) messages to all MSSs by the initiator MSS. $2*N_{min}*C_{air}$ is the total cost of sending checkpoint request message to the minimum number of processes that need to take checkpoints($N_{min}*C_{air}$) and reply to the initiator after taking the tentative checkpoint($N_{min}*C_{air}$). Hence algorithm [2] generates the global consistent state by using $N_{min}+ N_{indu}$ average number of checkpoint and $3*C_{broad} + 2*N_{min} * C_{air}$ message overhead cost but our proposed algorithm by using $N_{min}$ and $3* N_{min} * C_{air}$ respectively [Refer Table 1].. Thus algorithm [2] takes less useless checkpoint in the comparison of [1] but have high message overhead cost. The algorithm suffers from useless checkpoint and has higher message overhead as compared to the proposed algorithm.

The koo-Toueg[4] proposed a minimum process coordinated checkpointing algorithm for distributed systems with the cost of blocking of processes during checkpointing. This algorithm requires minimum number of synchronization message and number of checkpoint .In Toueg algorithm requires only minimum number of process to take checkpoints (ii) message overhead is $3*N_{min}*N_{dep} * C_{air}$ (iii) Blocking time is $N_{min}*T_{ch}$. Our proposed algorithm reduces the message overhead $3*N_{min}*N_{dep} * C_{air}$ *to* $3*N_{min}* C_{air}$ [Refer Table 1] without blocking of underlying processes.

In [8,25] authors designs an all process non blocking checkpointing algorithm. In these algorithms the initiator broadcast the checkpoint request to all processes the overhead of which is $C_{broad}$. The initiator receives reply from the N processes the overhead of which is $N*C_{air}$. At last the initiator broadcasts a commit request to all processes to convert their tentative checkpoints to permanent one. In such way we get the consistent global state with the total message overhead of $(2*C_{broad} + N*C_{air})$ [Refer Table 1]. However algorithm [8,25] had fewer messages overhead in the comparisons of our proposed algorithm but these algorithms forces to all processes in the system to take their checkpoints for each checkpoint initiation. This may waste the energy and processor power of the processes which are in doze mode. Compared to [8], our algorithm forces only a minimum number of processes to take checkpoint on stable storage.

### 6.3. A comparative Study:

In Elnozhay et al.[8] and S.Neogy et al.[25] algorithm proposed non blocking checkpointing algorithms but requires all-processes to take checkpoints during checkpointing, even though many of them may not be necessary. In mobile environment, since checkpoints need to be transferred to the stable storage at the MSSs over the wireless network. So in this way taking unnecessary checkpoints may waste a large amount of wireless bandwidth. In the algorithms [4,13] authors proposed minimum process checkpointing algorithm but it block its underlying computation during checkpointing. The blocking time of the Koo-Toueg[4] ($N_{min}* T_{ch}$)algorithm is highest, followed by Cao-Singhal[13] which is $2T_{st}$ (not shown in the table1). Therefore, blocking algorithm may degrade the performance to mobile computing systems [8]. The message overhead in proposed algorithm is greater than [8] but less than [1]. However, the algorithm in [8] is a centralized algorithm and there is no easy way to make it distributed without increasing message overhead. The Parkash-Singhal[15] proposed the first minimum process non blocking checkpointing algorithm. However author found that this algorithm may result in an inconsistency [3,13] in some situation and proved that there does not exist a non-blocking algorithm which forces only a minimum number of processes to take their checkpoints. Cao and Singhal [1] achieved non-intrusiveness in the minimum-process algorithm by introducing the concept of mutable checkpoints but number of useless checkpoints in [1] may be exceedingly high in some situations [19]. Also concurrent executions is allowed in [1], but in algorithm [20] author prove that algorithm [1] may lead to inconsistency during concurrent execution. Kumar et. al [2] proposed a five phase checkpointing algorithm to reduced the height of the checkpointing tree and the number of useless checkpoints by keeping non-intrusiveness intact. However, algorithm [2] reduces the useless checkpoint in the comparison of algorithm [1] but has extra message overhead cost. Our proposed coordinated checkpointing algorithm for mobile distributed system is non-blocking and forces only required minimum number of processes to take their checkpoints.

## 7. Conclusion

In this paper, we have designed a minimum process non-blocking coordinating checkpointing protocols which are suitable for mobile distributed environment. The main feature of algorithm are:*(1)* The number of processes that take checkpoints is minimized to avoid awakening of MHs in doze mode of operation.*(2)* no useless checkpoint are taken as [1,2]. *(3)* It is free from the avalanche effect. *(4)* Algorithm is non-blocking and not suspends their underlying computation during checkpointing.*(5)* save limited battery life of MHs and low bandwidth of wireless channels *(6)* reduces the latency associated with checkpoint request propagation compared to[4]. Thus our proposed algorithm has low communication and storage overheads. These all features make our proposed algorithm more suitable for mobile computing environment than the algorithms mentions in table 1.

---

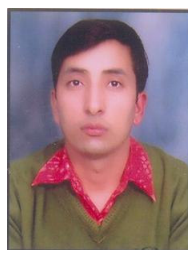

**Surender Kumar** is Sr. Lecturer in the Department of Information Technology at Haryana College of Technology & Management Kaithal (Haryana) India. He is pursuing his PhD in Computer Science from Kurukshetra University, Kurukshetra and his M.Tech. from the Ch. Devi Lal University, Sirsa(Haryana) INDIA. His research interests include checkpointing in mobile distributed systems, fault tolerance, mobile computing.

**Dr. R.K Chauhan** is serving as a Chairman, Department of Computer Sc. & application, Kurukshetra University, Kurukshetra. He has contributed many technical papers in areas including Database, Data Mining & Warehousing, Mobile Computing, Ad-hoc Networks and Checkpointing Algorithms.

**Dr. Parveen Kumar** received his Ph.D degree in Computer Sc. from Kurukshetra University, Kurukshetra, India, in 2006. He has contributed over 20 technical papers in areas including Checkpointing Algorithms, Mobile Computing and Distributed Systems. He is serving in MIET, Meerut(U.P), INDIA as a Professor in Computer Sc. & Engineering department.